# Oxygen reduction reactions on pure and nitrogen-doped graphene: a first-principles modeling

Danil W. Boukhvalov[1,*] and Young-Woo Son[1]

[1]School of Computational Sciences, Korea Institute for Advanced Study, Seoul 130-722, Korea.

Based on first principles density functional theory calculations we explored energetics of oxygen reduction reaction over pristine and nitrogen-doped graphene with different amounts of nitrogen doping. The process of oxygen reduction requires one more step then same reaction catalyzed by metals. Results of calculations evidence that for the case of light doped graphene (about 4% of nitrogen) energy barrier for each step is lower than for the same process on Pt surface. In contrast to the catalysis on metal surface the maximal coverage of doped graphene is lower and depends on the corrugation of graphene. Changes of the energy barriers caused by oxygen load and corrugation are also discussed.

Low temperature fuel cells attract significant attentions regarding possible production of electricity by the direct electrochemical conversation of hydrogen and oxygen to water.[1] Practical application of fuel cells has been hindered largely by the slow kinetic of oxygen reduction reactions (hereafter ORR). The platinum-based catalysts were proposed for the increasing efficiency and speed of ORR.[2-9] Several problems such as high cost and rarity of materials, diffusion of fuel molecules into the electrodes and CO poisoning[10] have been discussed for metal based catalysts.[2-9,11] Meanwhile, nitrogen doped carbon nanosystems were intensively studied in the last couple of years as possible alternative to metal-based catalysts.[12-27] Light weight, low cost and high efficiency makes these materials attractive for the further practical applications.

Further development of nitrogen doped carbon materials requires a proper theoretical description of all steps of ORR over these catalysts. Recent reports on catalytic activity of N-doped graphene[24-27] and development of the various methods of implementation of nitrogen in graphene[24-29] suggest us to start building of adequate modeling of this new kind ORR catalysis from nitrogen doped graphene. In previous theoretical studies were considered only charge redistribution near nitrogen impurity in single wall carbon nanotubes[23] or pieces of nanographenes.[28] Roles of the nitrogen on zigzag shaped edges of graphite for the oxygen adsorption and reduction were also considered.[30,31] The binding of oxygen peroxide with nitrogen doped nanographenes were also discussed theoretically.[27] The effect of oxygen coverage[32,33] and graphene corrugation,[34] however, on the adsorption and desorption processes have not been discussed so far although those effects are expect to occur naturally in the ORR process.

In this Communication we report the calculations of energetics for all steps of ORR over pristine graphene, nitrogen doped graphene and corrugated graphene at different levels of oxygen load and nitrogen amount for possible applications of graphene as catalysts. Owing to changes in molecular bonding characteristics between oxygen and doped graphene, the nitrogen doped graphene is shown to be the most promising candidate for catalytic applications. We also find the experimentally detected[8] doping level of 4-5% of nitrogen is proper for the applications. Therefore, the present model of ORR over pure and doped graphene could be used for description of results for other carbon based materials[12-22] and further prediction of catalytic effect of different amount of various dopants.

The modeling was performed by density functional theory (DFT) in the pseudopotential code SIESTA,[35] as was done in our previous works.[32-34,36] All calculations were performed using the generalized gradient approximation (GGA-PBE) with spin-polarization.[37] Full optimization of the atomic positions was performed. During the optimization, the ion cores are described by norm-conserving non-relativistic pseudo-potentials[38] with cut off radii 1.14, 1.45 and 1.25 a.u. for C, N and H respectively, and the wavefunctions are expanded with a double-ζ plus polarization basis of localized orbitals for carbon and oxygen, and a double-ζ basis for hydrogen. Optimization of the force and total energy was performed with an accuracy of 0.04 eV/Å and 1 meV, respectively. All calculations were carried out with an energy mesh cut-off of 360 Ry and a k-point mesh of 8×6×1 in the Mokhorst-Park scheme.[39] Further increase of energy mesh cut-off to 420 Ry and k-point mesh to 16×12×2 provide changes in total energy of system less than 10 meV. The calculation of the free energy were performed by previously developed method[8,9] for platinum based catalysis formula: $G = \Delta E_N - neU + E_{ZP}$, where $\Delta E_N$ is the energy difference between total energies at N and N−1 step of reaction processes, $e$ an electron charge,

U an equilibrium potential, $n$ the number of oxygen atoms, and $E_{ZP}$ the zero point energy correction, respectively. The values of U (1.23 eV)[40,41] and zero point energy corrections are same as ones used in previous works.[8,9] External energy field that can also change the chemisorption energy[42] was not included in our simulation. For our modeling we have used rectangle like supercell containing 48 carbon atoms. The modeling of corrugation has been done same way as in our previous work.[34] The supercell have been initially bended with the difference between the lowest and highest points about 1 Å. Further optimization has been performed with the keeping of this height of bend by the fixation of the atomic positions of four atoms at lowest and highest part of the graphene membrane.

First, we explore the ORR over pristine graphene substrate. Oxygen molecule forms a weak ionic bond to graphene and graphene-oxygen distance at the initial step of reaction is 2.56 Å (Fig. 1a). At the second step of reaction are created intermediate metastable configuration (Fig. 1c) when the oxygen atom is covalently bonded with carbon atom. The carbon-oxygen distance at this stage is 1.44 Å which is bigger than standard C-O value (about 1.2 Å). In contrast to the case of metal based catalysis when oxygen atoms bonded only with atoms of metal surface, this step of ORR over graphene is endothermic (see Fig. 2a). Third step of graphene-based ORR is formation of epoxy groups on graphene surface (Fig. 1e). This intermediate configuration does not form over metallic surface. Transition from previous metastable configuration to the stable epoxy groups is significantly energetically favorable[32,43] (Fig. 2a), and the next step of oxygen reduction – formation of hydroxyl groups (Fig. 1g) is also exothermic in contrast to the reduction over Pt(111) surface (see Fig. 2a and Ref. [3]). Last step (Fig. 1i) of ORR is the same as for metallic catalysts. From our calculation results, we can conclude that the characteristics of carbon-oxygen bonds provides different energetics of ORR if compared to the case of metallic scaffolds and that pure graphene is unsuitable for catalysis because of rather high energy cost for the second step of reaction (Fig. 1c).

Unlike the pristine graphene, the nitrogen-doped graphene is found to have a dramatic lower energy cost for the early catalytic reaction steps. Our present findings are also consistent with the previous work[37,42] suggesting higher chemical activity of nitrogen impurities in graphene. We examine three levels of nitrogen doping (low (one nitrogen atom per used supercell – 2% of nitrogen), medium (the pair of nitrogen atoms – 4% of nitrogen) and highest (half of carbon atoms were replaced by nitrogen). For the case of 4% N-doped graphene the most energetically favorable configuration is determined by changing relative positions between nitrogen impurities (Fig. 1b). The obtained optimized configuration is further used for other catalytic reactions. As shown in Fig. 1b and d, initial steps of the oxygen adsorption near nitrogen impurity is drastically different from the case of pristine graphene (Fig. 1a, c). At the second step of reaction the metastable configuration discussed for pristine graphene does not form. Oxygen forms ionic bond with N-doped graphene with charge transfer of 0.86$e$ from graphene (here $e$ is an electron charge). In this case, the distance between graphene and oxygen is about 2.30 Å which is much smaller than van der Waals radius of 3.5 Å. The energy cost of this step is much smaller than one for pristine graphene. Last steps of ORR over doped graphene are similar to the pristine graphene and metallic surfaces. For the case of significant nitrogen doping, we have obtained the irreversible oxidation of graphene with significant (more than 3eV) desorption barriers. From the calculation so far, we can conclude that the low amount of nitrogen doping (closer to experimentally reported values of 5%)[24] enables ORR to occur on graphene scaffold with much smaller initial energy costs.

Realistic description of the chemical reactions required also estimation of the values of the energy barriers for the intermediate steps of reactions. For intermediate steps between 1st and seconds and 3rd and 5th steps of reaction the energy barriers is the same as for the case of remote molecules and atoms migration to graphene surface and does not exceed 0.1 eV.[44] The value of the energy barrier between 2nd and 3rd steps of reaction is corresponding with the activation of oxygen on graphene.[43] We have calculated this values for the different amount of nitrogen and found that its decrease from 1.09 eV for pure graphene (this value is near to reported before 1.04 eV for nanographenes[43]), throughout 0.67 eV for 2% of nitrogen to 0.19 eV for the case of 4% of nitrogen. Obtained values of the energy barriers for the intermediate points between steps of reactions over 4% N-doped graphene does not change significantly the energetics of reduction process.

In contrast to the stiff metallic surface graphene is flexible and corrugation caused by the chemisorption could drastically change the energetics of the process.[31,33] To incorporate this effect in graphene catalyzed ORR we step by step increase the number of oxygen atoms on 4% N-doped graphene. Results of our calculations (see Fig. 2b) suggest that increase of coverage from 4 to 16% (Fig. 1k) do not change the energetics of reaction. Further increasing of the oxygen coverage provides spontaneous desorption of hydroxyl groups. This results is in good agreement with vanishing of hydroxyl groups from graphene oxide after nitrogen doping.[25,27]

Experimental observation indicates that N-doped graphene in catalytic processes is wrinkled.[24] For the modeling of this realistic graphene we performed calculations for the bended graphene (see side view on Fig. 1l). Results of the calculation of the free energy show that for the lowest oxygen coverage the energy barriers became smaller than those for the flat graphene, but increase of oxygen amount leads increase of energy costs for intermediate steps of ORR (see Fig. 2c).

In summary, based on our first principles modeling, we demonstrate that the intermediate steps of ORR over graphene scaffold are drastically different from ones on metallic substrate. Doping of graphene by small amount of nitrogen leads significant decrease in the energy costs at the intermediate steps of oxygen reduction. Thus, N-doped graphene is promising for catalysts of oxygen reduction. The free energy of ORR over N-doped graphene weakly depends on the oxygen coverage when it is less than 16 %. We also demonstrate that the free energy of ORR increases significantly for the distorted graphene.

ACKNOWLEDGMENT

We acknowledge computational support from the CAC of KIAS. Y.-W. S. was supported by the NRF grant funded by MEST (Quantum Metamaterials Research Center, R11-2008-053-01002-0 and Nano R&D program 2008-03670).

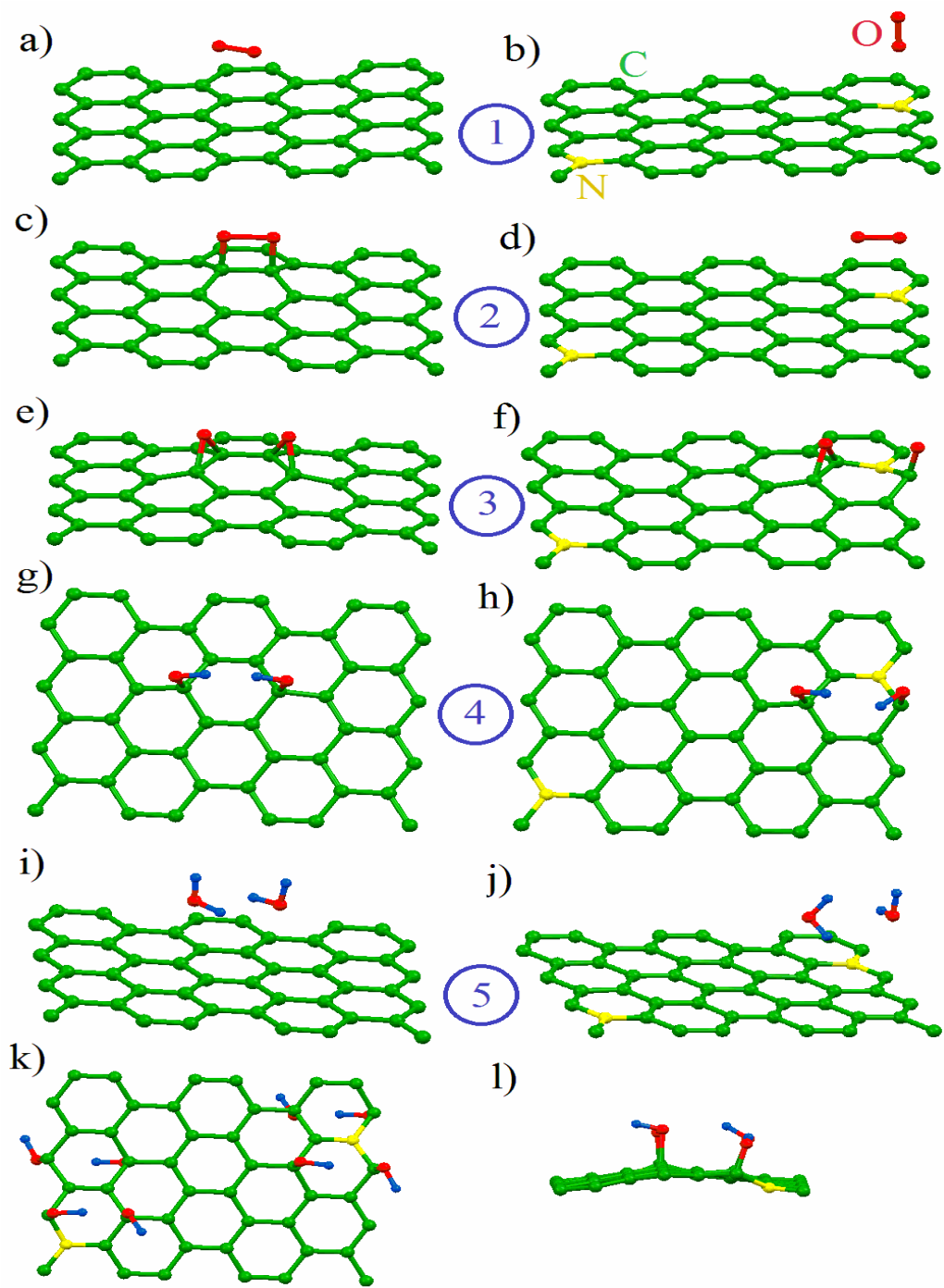

**Figure 1** Optimized atomic structure for the five steps (shown by the numbers in circles) of ORR on pure (a, c, e, g, i) and 4% N-doped (b, d, f, h, j) graphene scaffold, and fourth step of oxygen reduction on 4% N-doped flat (k) and corrugated (l) graphene for the 16 (k) and 8 (l) surface coverage.

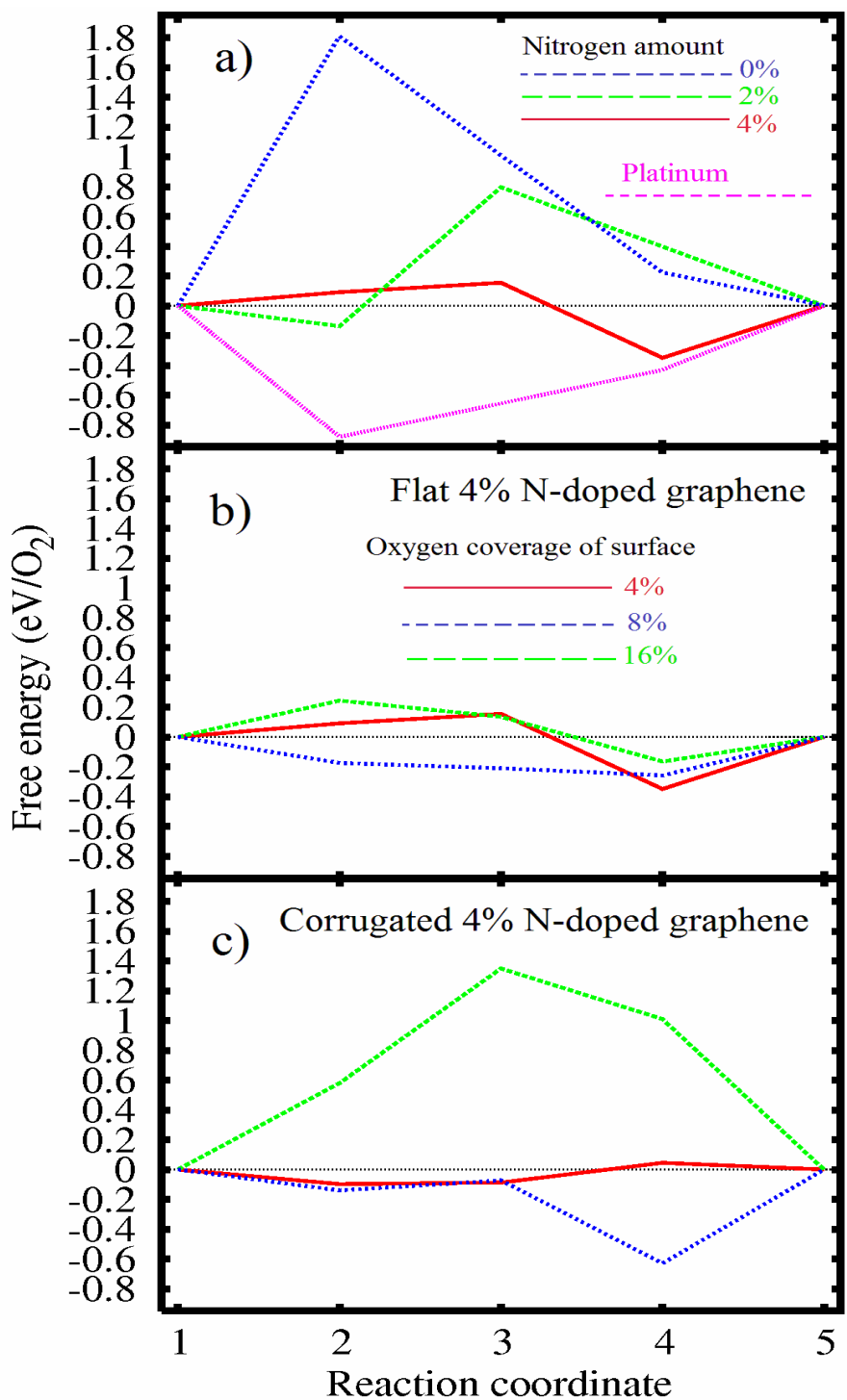

**Figure 2** Free energy diagram for ORR over pure and doped graphene for lowest surface coverage (4%) by oxygen and Pt(111) surface from Ref. [3] (a), for different oxygen's coverage of 4% N-doped flat (b) and corrugated (c) graphene.